\documentclass[11pt]{article}
\usepackage{geometry}                
\usepackage{graphicx}
\usepackage{amssymb}
\usepackage{epstopdf}
\usepackage{amsmath}

\title{Schr\"odinger-Pauli Equation for the Standard Model Extension CPT-Violating Dirac Equation
}
\author{Thomas D. Gutierrez\\Physics Department\\California Polytechnic State University, San Luis Obispo}
\date{\today}                                           

\begin{document}
\maketitle
It is instructive to investigate the non-relativistic limit of the simplest Standard Model Extension (SME) CPT-violating Dirac-like equation \cite{colladay:1997} but with minimal coupling to the electromagnetic fields.  In this limit, it becomes an intuitive Schr\"odinger-Pauli-like equation.   This is comparable to the free particle treatment as explored by Kostelecky and Lane \cite{lane:1999}, but this exercise only considers the $a$ and $b$ CPT-violating terms and $\vec{p}/m$ terms to first order.  Several toy systems are discussed.

\section*{Schr\"odinger-Pauli Equation}
The simplest free particle SME CPT-Violating Dirac-like equation is given by Eq.~6 in \cite{colladay:1997} (equivalent to the equations of motion using the Lagrangian from Eq.~1 in \cite{lane:1999} with SME parameters $c=d=e=f=g=H=0$)
\begin{equation}
(i\gamma^\mu\partial_\mu-a_\mu\gamma^\mu-b_\mu\gamma_5\gamma^\mu-m\mathbb{I})\Psi=0
\label{SME}
\end{equation}
where the parameters $a_\mu$ and $b_\mu$ represent the CPT-violating constants written as 4-vector-like quantities.  In the limit where $a_\mu=b_\mu=0$, the familiar covariant form of the Dirac equation is recovered

\begin{equation}
(i\gamma^\mu\partial_\mu-m\mathbb{I})\Psi=0.
\label{cDirac}
\end{equation}
To obtain the Schr\"odinger-Pauli equation for Eq.~(\ref{SME}), it is convenient to review a simple method for obtaining the Schr\"odinger-Pauli equation for the ordinary Dirac equation.  To do this, it is useful to first massage Eq.~(\ref{cDirac}) before introducing the minimal coupling to the electromagnetic field.   Using natural units throughout, the following conventions for the various $4\times 4$ matrices are used in the calculation:

\begin{equation}
\gamma^\mu=(\beta,\beta\vec{\alpha})
\label{gamma}
\end{equation}

with

\begin{equation}
\beta=\left(\begin{array}{cc}
I & 0 \\
0 & -I  \end{array} \right),~~
\vec{\alpha}=\left( \begin{array}{cc}
0 & \vec{\sigma} \\
\vec{\sigma} & 0  \end{array} \right),~~
\gamma_5=\left( \begin{array}{cc}
0 & I \\
I & 0  \end{array} \right),~~
\mathbb{I}=\left( \begin{array}{cc}
I & 0 \\
0 & I  \end{array} \right)
\label{matrices}
\end{equation}
where $\vec{\sigma}$ are the usual $2\times 2$ Pauli-matrices and $I$ represents the $2\times 2$ unit matrix.  The relevant kinematic quantum operators are given by

\begin{equation}
i\partial_\mu=i(\frac{\partial}{\partial t},\nabla)=p_\mu\equiv(E,-\vec{p}),
\label{ops}
\end{equation}
and the (time-dependent) wave function is a Dirac bispinor

\begin{equation}
\Psi(\vec{r},t)=\left( \begin{array}{c}
\Psi_A  \\
\Psi_B  \\
\end{array} \right), 
\end{equation}
where $\Psi_A$ is the ``upper'' spinor and $\Psi_B$ is the ``lower'' spinor.  Now, substituting the matrices of Eq.~(\ref{matrices}) into the gamma matrix equations in Eq.~({\ref{gamma}}) and that, along with the operators in Eq.~(\ref{ops}), all into Eq.~(\ref{cDirac}) gives the following familiar form of the Dirac equation in terms of the upper and lower wave functions
\begin{align}
(\vec{\sigma}\cdot\vec{p})~\Psi_B&=(E-m)~\Psi_A \label{diracup}\\
(\vec{\sigma}\cdot\vec{p})~\Psi_A&=(E+m)~\Psi_B.\label{diraclo}
\end{align}

The following calculation basically follows the reasoning used in Halzen and Martin (HM) exercise 5.5 leading to Schr\"odinger-Pauli equation (equation 5.31 in HM).  

First,  the wave function is minimally coupled to the electromagnetic scalar and vector potentials, $\phi$ and $\vec{A}$,  with the the substitution of the canonical momentum-energy operators into Eqns.~(\ref{diracup}) and (\ref{diraclo})
\begin{equation}
p^\mu\rightarrow p^\mu-qA^\mu
\end{equation}
or, more explicitly, with $A^\mu=(\phi,\vec{A})$,
\begin{equation}
E\rightarrow~E-q\phi,~~\vec{p}\rightarrow~\vec{p}-q\vec{A}
\label{canon}
\end{equation}
leading to 
\begin{align}
\vec{\sigma}\cdot(\vec{p}-q\vec{A})~\psi_B&=(E-q\phi-m)~\psi_A \label{up}\\
\vec{\sigma}\cdot(\vec{p}-q\vec{A})~\psi_A&=(E-q\phi+m)~\psi_B \label{lo}.
\end{align}
Note the shift in notation in Eqs.~(\ref{up}) and (\ref{lo}).  With $E$ taken as an operator in Eqs.~(\ref{diracup}) and (\ref{diraclo}), the usual phasor solutions to $\Psi$ in the time domain emerge after separation of variables.  Since, in the spirit of the Schr\"odinger-Pauli Equation, we seek energy eigenvalues, the parameter $E$ in Eq.~(\ref{sp1}) is now taken as the energy eigenvalues of the Hamiltonian rather than an operator; the wave functions, $\psi$, are now only a function of position.

From Eq.~(\ref{lo}), solve for $\psi_B$ giving
\begin{equation}
\psi_B=\frac{\vec{\sigma}\cdot(\vec{p}-q\vec{A})}{(E-q\phi+m)}~\psi_A.
\label{psiB}
\end{equation}
In the non-relativistic limit, the ``non-relativistic energy'' is defined as $E_{\rm NR}=E-m$ where it is understood that $m\gg E_{\rm NR}$, $E\sim m$, and we take $m\gg q\phi$.  In this limit, from Eq.~(\ref{psiB}), $\psi_B$ will be small compared to $\psi_A$.  Substituting $\psi_B$ into Eq.~({\ref{up}) and simplifying gives
\begin{equation}
\left[\vec{\sigma}\cdot(\vec{p}-q\vec{A})\right]\left[\vec{\sigma}\cdot(\vec{p}-q\vec{A})\right]~\psi_A=(E-q\phi+m)(E-q\phi-m)~\psi_A.
\label{sp1}
\end{equation}
The right hand side of Eq.~(\ref{sp1}) can be approximated in the non-relativistic limit to be

\begin{equation}
(E-q\phi+m)(E-q\phi-m)~\psi_A\sim 2m(E_{\rm NR}-q\phi)~\psi_A.
\label{en}
\end{equation}

The left hand side of Eq.~(\ref{sp1}) is more subtle because it involves the two operators $\vec{p}=-i\nabla$ and $\vec{A}$ that don't commute (since $\vec{A}$ is generally a function of position).  To simplify, use the vector identities
\begin{equation}
\nabla{(f\vec{Q})}=f\nabla\times\vec{Q}-\vec{Q}\times\nabla f
\label{id1}
\end{equation}
and the Pauli spin matrix identity
\begin{equation}
(\vec{\sigma}\cdot\vec{Q})(\vec{\sigma}\cdot\vec{R})=(\vec{Q}\cdot\vec{R})I+i\vec{\sigma}\cdot(\vec{Q}\times\vec{R})
\label{id2}
\end{equation}
with $\vec{Q}=\vec{R}=(\vec{p}-q\vec{A})$.  The elements of the right hand side of Eq.~(\ref{id2}) become 

\begin{equation}
\vec{Q}\cdot\vec{R}=(\vec{p}-q\vec{A})^2
\end{equation}
and
\begin{equation}
\vec{Q}\times\vec{R}=-q(\vec{p}\times\vec{A}+\vec{A}\times\vec{p})\label{cross}.
\end{equation}
To simplify the right hand side of Eq.~(\ref{cross}), use it (as an operator) on a trial function $\psi_A$ along with the identity in Eq.~(\ref{id1}) with $f=-i\psi_A$ and $\vec{Q}=\vec{A}$ so Eq.~(\ref{cross}) becomes
\begin{equation}
\vec{Q}\times\vec{R}=+iq\nabla\times\vec{A}=+iq\vec{B}
\end{equation}
where the magnetic field is $\vec{B}=\nabla\times\vec{A}$.  Putting all this together into Eq.~(\ref{sp1}) along with the approximation from Eq.~(\ref{en}) and simplifying gives
\begin{equation}
\left[\frac{1}{2m}(\vec{p}-q\vec{A})^2-\frac{q}{2m}(\vec{\sigma}\cdot\vec{B})+q\phi\right]\psi_A=E_{\rm NR}~\psi_A,
\label{sp}
\end{equation}
which is the Schr\"odinger-Pauli equation.  The advantage to this form is that it very clearly highlights, in a familiar single-particle first-quantized equation, the electric and magnetic contributions to the wave equation and how they interplay with spin coming from the limit of a relativistically responsible source (i.e. the covariant Dirac equation).  


\section*{Schr\"odinger-Pauli Equation for SME Dirac-like Equation}
Now the same recipe as described above for the ordinary Dirac equation is performed for Eq.~(\ref{SME}), the SME Dirac equation. Before minimal coupling, the matrices of Eqs.~(\ref{matrices}) and ({\ref{gamma}}) are substituted into into Eq.~(\ref{SME}), along with the operators in Eq.~(\ref{ops}).  The CPT-violating terms are new elements worth writing explicitly

\begin{align}
a_\mu\gamma^\mu=& ~\beta a_0-\beta\vec{a}\cdot\vec{\alpha} \\
b_\mu\gamma_5\gamma^\mu=& ~\beta\gamma_5 b_0-\beta\gamma_5\vec{b}\cdot\vec{\alpha}.
\end{align}
With these substations (and exploiting $\beta^2=\mathbb{I}$), Eq.~(\ref{SME}) can be written

\begin{equation}
(\vec{\alpha}\cdot\vec{p}+\beta m+a_0-\vec{\alpha}\cdot\vec{a}+\gamma_5 b_0-\gamma_5 \vec{\alpha}\cdot\vec{a})\psi=E\psi
\label{SME2}
\end{equation}
where we have transitioned to the time-independent form so $\vec{p}$ is understood to be an operator and $E$ is an eigenvalue.  Further expanding and simplifying Eq.~(\ref{SME2}) gives the following coupled equations between the upper and lower components of $\psi$
\begin{align}
(\vec{\sigma}\cdot\vec{p}+b_0-\vec{\sigma}\cdot\vec{a})~\psi_B&=(E-m-a_0+\vec{\sigma}\cdot\vec{b})~\psi_A \label{smeup}\\
(\vec{\sigma}\cdot\vec{p}+b_0-\vec{\sigma}\cdot\vec{a})~\psi_A&=(E+m-a_0+\vec{\sigma}\cdot\vec{b})~\psi_B.\label{smelo}
\end{align}
Now, solving for $\psi_B$ using Eq.~(\ref{smelo}),
\begin{equation}
\psi_B=\frac{\vec{\sigma}\cdot\vec{p}+b_0-\vec{\sigma}\cdot\vec{a}}{(E+m-a_0+\vec{\sigma}\cdot\vec{b})}~\psi_A
\label{smepsiB}
\end{equation}
and substitute it into Eq.~(\ref{smeup}) giving

\begin{equation}
(\vec{\sigma}\cdot\vec{p}+b_0-\vec{\sigma}\cdot\vec{a})^2\psi_A=(E-m-a_0+\vec{\sigma}\cdot\vec{b})(E+m-a_0+\vec{\sigma}\cdot\vec{b})\psi_A.
\label{smepsiC}
\end{equation}
The left hand side of Eq.~(\ref{smepsiC}) can then be expanded into
\begin{equation}
(\vec{\sigma}\cdot\vec{p})(\vec{\sigma}\cdot\vec{p})-(\vec{\sigma}\cdot\vec{p})(\vec{\sigma}\cdot\vec{a})-(\vec{\sigma}\cdot\vec{a})(\vec{\sigma}\cdot\vec{p})+2b_0(\vec{\sigma}\cdot\vec{p})+(\vec{\sigma}\cdot\vec{a})(\vec{\sigma}\cdot\vec{a})-2b_0(\vec{\sigma}\cdot\vec{a})+b_0^2,
\label{lhsC}
\end{equation}
which preserves the order of the relevant $\sigma$ and $\vec{p}$ products in anticipation of minimal coupling where, as before, the commutativity of the terms must be considered.  However, the $a$ and $b$ terms are still considered constants.  It is also helpful to begin making approximations.  The various $a$ and $b$ constants are presumably small, so Eq.~(\ref{lhsC}) can be simplified by only keeping leading order terms in $a$ and $b$
\begin{equation}
(\vec{\sigma}\cdot\vec{p})(\vec{\sigma}\cdot\vec{p})-(\vec{\sigma}\cdot\vec{p})(\vec{\sigma}\cdot\vec{a})+(\vec{\sigma}\cdot\vec{a})(\vec{\sigma}\cdot\vec{p})+2b_0(\vec{\sigma}\cdot\vec{p}).
\label{lhsC2}
\end{equation}
After implementing the minimal coupling as prescribed by Eq.~(\ref{canon}), each term can be considered in turn.  As before,

\begin{equation}
(\vec{\sigma}\cdot\vec{p})(\vec{\sigma}\cdot\vec{p})\rightarrow(\vec{p}-q\vec{A})^2-q(\vec{\sigma}\cdot\vec{B})
\end{equation}
and now
\begin{equation}
2b_0(\vec{\sigma}\cdot\vec{p})\rightarrow2b_0[\vec{\sigma}\cdot(\vec{p}-q\vec{A})].
\end{equation}
Using the identity in Eq.~(\ref{id2}), the other terms can be simplified so
\begin{equation}
(\vec{\sigma}\cdot\vec{p})(\vec{\sigma}\cdot\vec{a})=(\vec{p}\cdot\vec{a})I+i(\vec{p}\times\vec{a})\cdot\vec{\sigma}
\end{equation}
and
\begin{equation}
(\vec{\sigma}\cdot\vec{a})(\vec{\sigma}\cdot\vec{p})=(\vec{a}\cdot\vec{p})I+i(\vec{a}\times\vec{p})\cdot\vec{\sigma}=(\vec{p}\cdot\vec{a})I-i(\vec{p}\times\vec{a})\cdot\vec{\sigma}.
\end{equation}
Now, with Eq.~(\ref{canon}),
\begin{equation}
-(\vec{\sigma}\cdot\vec{p})(\vec{\sigma}\cdot\vec{a})-(\vec{\sigma}\cdot\vec{a})(\vec{\sigma}\cdot\vec{p})\rightarrow-2[\vec{a}\cdot(\vec{p}-q\vec{A})]I.
\end{equation}
Putting it all together, the left hand side of Eq.~(\ref{SME2}) can be written
\begin{equation}
[(\vec{p}-q\vec{A})^2-q(\vec{\sigma}\cdot\vec{B})-2\vec{a}\cdot(\vec{p}-q\vec{A})+2b_0\vec{\sigma}\cdot(\vec{p}-q\vec{A})]\psi_A
\label{smel}
\end{equation}
where the $I$ matrices have been suppressed.  

For the right hand side of Eq.~(\ref{smepsiC}), after applying Eq.~(\ref{canon}), a non-relativistic approximation is made where $E_{\rm NR}=E-m$, $m\gg E_{\rm NR}$, $E\sim m$, and $m\gg q\phi$.  Again, only leading order terms in $a$ and $b$ are kept.  In this limit, the right hand side becomes
\begin{equation}
(E-q\phi-m-a_0+\vec{\sigma}\cdot\vec{b})(E-q\phi+m-a_0+\vec{\sigma}\cdot\vec{b})~\psi_A\sim 2m(E_{\rm NR}-q\phi-a_0+\vec{\sigma}\cdot\vec{b})~\psi_A.
\label{smer}
\end{equation} 
Setting Eq.~(\ref{smel}) and Eq.~(\ref{smer}) equal, and then tidying (and restoring $\hbar$ and $c$ into unnatural units), gives the Schr\"odinger-Pauli Equation for the SME Dirac-like equation:
\begin{equation}
\left[\underbrace{\frac{(\vec{p}-q\vec{A})^2}{2m}-\frac{q\hbar}{2m}(\vec{\sigma}\cdot\vec{B})+q\phi}_\text{SP Hamiltonian from Eq.~(\ref{sp})}\underbrace{+a_0-\frac{\vec{a}\cdot(\vec{p}-q\vec{A})}{m}-\hbar\vec{\sigma}\cdot\vec{b}+\frac{b_0\hbar\vec{\sigma}\cdot(\vec{p}-q\vec{A})}{m}}_\text{SME CPT-violating terms}\right]\psi_A=E_{\rm NR}~\psi_A.
\label{spsme}
\end{equation} 
With $\vec{A}=\vec{B}=\phi=0$, the operator on the left hand side of Eq.~(\ref{spsme}) is equivalent to the non-relativistic Hamiltonian in \cite{lane:1999} (their Eq. 26) with SME parameters $c=d=e=f=g=H=0$ (and taken only to first order in $\vec{p}/m$ beyond the kinetic term).  

Equation (\ref{spsme}) can be written in an illuminating way that highlights the features of the CPT-violating terms
\begin{equation}
\left[\frac{(\vec{p}-q\vec{A})^2}{2m}-\frac{q\hbar}{2m}(\vec{\sigma}\cdot\vec{\mathcal{B}})+q{\mathcal{S}}\right]\psi_A=E_{\rm NR}~\psi_A,
\label{spsme2}
\end{equation}
where
\begin{equation}
\vec{\mathcal{B}}=\vec{B}+\frac{2m}{q}\vec{b}-\frac{2b_0}{q}(\vec{p}-q\vec{A})\label{bb}
\end{equation}
and
\begin{equation}
\mathcal{S}= \phi+\frac{a_0}{q}-\frac{\vec{a}\cdot(\vec{p}-q\vec{A})}{mq}\label{pp}
\end{equation}
are a ``modified'' or effective magnetic field and scalar potential respectively.  This intuitive Schr\"odinger-Pauli form highlights a number of properties of the $a$ and $b$ CPT-violating parameters.  For example, $a_0$ merely acts as a constant offset to the scalar potential energy, so is not measurable in this model.  The $\vec{b}$ term acts as an intrinsic background magnetic field provided by the vacuum itself (e.g. even in the absence of any applied field). In addition, there are curious ``kinematic magnetic'' and ``kinematic scalar'' terms that couple, via $b_0$ and $\vec{a}$, directly to the canonical momentum vector operator to first order.  Here, $a_0$ will have units of energy, while $\vec{a}$ will have units of momentum.  So, in units of energy, $a^\mu=(a_0,\vec{a}c)$. Also, $b_0$ is units of wave number and $\vec{b}$ in units of angular frequency.  In units of energy, $\hbar b^\mu=(\hbar c b_0,\hbar\vec{b})$. 


\section*{Toy Examples}
The Schr\"odinger-Pauli form in Eq.~(\ref{spsme2}) lends itself to ordinary Schr\"odinger mechanics intuition.  In this context, it is worth considering a few toy textbook examples.  Below, $\psi_A$ and $E_{\rm NR}$ above are replaced with $\psi$ and $E$ as well as $\vec{p}=-i\hbar\nabla$ (where it is understood $\psi'(x)$ is the derivative with respect to $x$ etc.).

\subsection*{Free Particle with Non-Zero $\vec{a}$}
With a free spin-up particle, consider a universe where $a_0=a_y=a_z=b_0=\vec{b}=0$ but where $a_x\ne 0$. Putting these values into Eq.~(\ref{spsme2}) leads to an equation of motion

\begin{equation}
\psi''(x)-\frac{2ia_x}{\hbar}\psi'(x)=-\frac{2mE}{\hbar^2}\psi(x)
\end{equation}
and solutions of the form

\begin{equation}
\psi(x)=N e^{i (a_x x/\hbar)}e^{\pm ikx}
\label{fp}
\end{equation}
where

\begin{equation}
k=\frac{\sqrt{2mE+a_x^2}}{\hbar},
\end{equation}
leading to a shift in the free particle's momentum of order $a_x$ (regardless of the direction of propagation).  That is, $a_x$ can be interpreted as a ``wind'' pointing in a fixed direction in space that gives the particle a little kick if it is moving downwind and pushes back on it if it is heading upwind.   The existence of such a preferred orientation in free space has a clear Lorentz-violating character to it.

\subsection*{Particle in a Box with Non-Zero $\vec{a}$}
Place a spin-up particle in an infinite well of width $L$ ($q\phi=0$ for $0<x<L$ and infinite otherwise), again where $a_0=a_y=a_z=b_0=\vec{b}=0$ and where $a_x\ne 0$ and $\vec{A}=\vec{B}=0$.  Using the free particle solutions in Eq.~(\ref{fp}), and after imposing boundary conditions, the eigenfunctions are of the form

\begin{equation}
\psi(x)=N e^{i (a_x x/\hbar)}\sin{(\frac{n\pi x}{L})}
\end{equation}
with $n=1,2,3,...$. Other than a phase consistent with the free particle momentum shift, these are the same solutions as the standard particle in a box. The energy eigenvalues are

\begin{equation}
E_n=\frac{\hbar^2\pi^2}{2mL^2}n^2-\frac{a_x^2}{2m},
\end{equation}
which leads to an (unmeasurable) constant shift in the usual infinite square well energies.

\subsection*{One Dimensional Tunneling with Non-Zero $\vec{a}$}
A free spin-up particle with energy $E$ impinges on a barrier $q\phi=V_0$ for $0<x<L$ and zero otherwise.  Assume $E<V_0$ so $(V_0-E) >0$.  As before, consider the case where $a_0=a_y=a_z=b_0=\vec{b}=0$ and where $a_x\ne 0$ and $\vec{A}=\vec{B}=0$.  In the simplest treatment, the solutions outside the barrier are incident, reflected, and transmitted free particle solutions consistent with Eq.~(\ref{fp}) with the appropriate coefficients.  In the barrier, the equations of motion are of the form

\begin{equation}
\psi''(x)-\frac{2ia_x}{\hbar}\psi'(x)=\frac{2m(V_0-E)}{\hbar^2}\psi(x)
\end{equation}
with solutions

\begin{equation}
\psi(x)=N e^{i (a_x x/\hbar)}e^{\pm\kappa x}
\label{fpx}
\end{equation}
where 
\begin{equation}
\kappa=\frac{\sqrt{2m(V_0-E)-a_x^2}}{\hbar}.
\end{equation}
A conspicuous feature of this tunneling amplitude is if 

\begin{equation}
a_x^2>2m(V_0-E)
\end{equation}
then $\kappa$ becomes imaginary and the solutions in the barrier, Eq.~(\ref{fpx}), become {\em free particle} solutions.  That is, the $a_x$ term effectively lowers the height of the barrier (or increases the kinetic energy) and facilitates ``true tunneling'' with complete barrier transparency.  

\subsection*{Transverse Motion of a Free Particle with Non-Zero $b_0$}
Another system to consider is a free particle that is moving transverse to the orientation of its spin.  For example, if a particle is initially in a spin-up orientation along the $z$-direction and then is allowed to move as a free particle only in the $x$-direction.  For simplicity, consider $a_0=\vec{a}=\vec{b}=0$ and $\vec{A}=\vec{B}=\phi=0$, but where $b_0\ne 0$.  Plugging this into Eq.~(\ref{spsme}) gives, in three dimensions,

\begin{equation}
\left[\frac{-\hbar^2}{2m}\nabla^2-\frac{i b_0\hbar}{m}(\vec{\sigma}\cdot\nabla)\right]\psi=E~\psi,
\label{spsmeb3d}
\end{equation}
which reduces to 
\begin{equation}
\frac{-\hbar^2}{2m}\psi''(x)-\frac{i b_0\hbar}{m}\sigma_x\psi'(x)=E~\psi
\label{spsmeb1d}
\end{equation}
with the normalized spinor wave function
\begin{equation}
\psi(x)=\left( \begin{array}{c}
f(x)  \\
g(x)  \\
\end{array} \right).
\end{equation}
This leads to coupled differential equations in $f$ and $g$
\begin{align}
f''+\frac{2ib_0}{\hbar}g'(x)=-\frac{2mE}{\hbar^2}f \label{fg1}\\
g''+\frac{2ib_0}{\hbar}f'(x)=-\frac{2mE}{\hbar^2}g \label{fg2}.
\end{align}
If $b_0=0$, the spin-up and spin-down components decouple and the particle propagates in the $x$-direction consistent with its fixed initial spin conditions.  However, if $b_0\ne0$, for constant $E$, the result is inexorable spin-position correlations as the spin orientation changes depending on where the particle is measured in space.  

\section*{Conclusion}
The Schr\"odinger-Pauli form of the simplest CPT-violating Dirac equation, as derived in this paper, is ripe for both pedagogical and physical insight into SME models.  Several toy systems were explored, highlighting the functionality of this intuitive approach.

\section*{Acknowledgements}
Thanks to Matt Mewes for helpful discussions.
%
%
%
%
%
%



\end{document}